\documentclass[11pt, a4paper]{article}

\usepackage{algorithm}
\usepackage{algorithmic}
\usepackage{amsmath}
\usepackage{amssymb}
\usepackage{amstext}
\usepackage{authblk}
\usepackage[english]{babel}
\usepackage{bm}% bold math
\usepackage{braket}
\usepackage[labelsep=period]{caption}
\usepackage{dcolumn}% Align table columns on decimal point
\usepackage{dsfont}
\usepackage{lipsum}
\usepackage[margin=3.6cm]{geometry}
\usepackage{graphicx}
\usepackage{hyperref}
\usepackage{microtype}
\usepackage[onehalfspacing]{setspace}
\usepackage{subcaption}
\usepackage{url}

\def\rmd{\mathrm{d}}
\def\vc#1{\boldsymbol{\mathrm{#1}}}
\def\mt#1{\mathrm{#1}}
\def\T{^{\!\top}}

\title{Quantum-assisted Finite-element Design Optimization}
\author[1, 3   ]{Dyon van Vreumingen}
\author[1, 3, *]{Florian Neukart}
\author[1      ]{David Von Dollen}
\author[2      ]{Carsten Othmer}
\author[2      ]{Michael Hartmann}
\author[2      ]{Arne-Christian Voigt}
\author[3      ]{Thomas B\"ack}
\affil[1]{Volkswagen Group Region Americas, USA}
\affil[2]{Volkswagen Group Research, Germany}
\affil[3]{Leiden Institute for Advanced Computer Science, Leiden University, The Netherlands}
\affil[*]{\rm Corresponding author: Florian Neukart (florian.neukart@vw.com)}
\usepackage{cite}
\usepackage{enumitem}

\begin{document}

\maketitle

\begin{abstract}
Quantum annealing devices such as the ones produced by D-Wave systems are typically used for solving optimization and sampling tasks \cite{Benedetti2016, Smelyanskiy2017, Venturelli2015a, Jiang2017, Isakov2016, OGorman2015, Rieffel2014, Venturelli2015, Perdomo-Ortiz2015, Boixo2014, Babbush2014, Smolin2014, Perdomo-Ortiz2012, LosAlamos2016, Neukart2017b}, and in both academia and industry the characterization of their usefulness is subject to active research. Any problem that can naturally be described as a weighted, undirected graph may be a particularly interesting candidate \cite{Neukart2018, Neukart2017}, since such a problem may be formulated a as quadratic unconstrained binary optimization (QUBO) instance, which is solvable on D-Wave's Chimera graph architecture. In this paper, we introduce a quantum-assisted finite-element method for design optimization. We show that we can minimize a shape-specific quantity, in our case a ray approximation of sound pressure at a specific position around an object, by manipulating the shape of this object. Our algorithm belongs to the class of quantum-assisted algorithms, as the optimization task runs iteratively on a D-Wave 2000Q quantum processing unit (QPU), whereby the evaluation and interpretation of the results happens classically. Our first and foremost aim is to explain how to represent and solve parts of these problems with the help of a QPU, and not to prove supremacy over existing classical finite-element algorithms for design optimization.\\[4mm]

%\tiny\keyFont{\section{Keywords:} quantum computing, quantum physics, finite-element, design optimization, QUBO}
\end{abstract}
\newpage
\section{Introduction}
According to the laws of quantum mechanics, a quantum mechanical system, which is in the ground state (state of minimal energy) of a time-independent system, also remains in the ground state if a change to it happens only slowly, i.e. adiabatically. This is known as the adiabatic theorem. The idea of adiabatic quantum computing is to construct a system having a ground state that is still unknown at that time, which corresponds to solving a particular problem, and another one whose ground state is easy to prepare experimentally. Subsequently, the easy-to-prepare system is adiabatically transferred to the system whose ground state one is interested in, and then measured. If the transition is slow enough, one can obtain a minimum-energy solution to the problem. D-Wave's QPUs deploy a system described by the two-dimensional Ising spin hamiltonian \cite{Neukart2018, Neukart2017}:
\begin{equation}
\label{equation1}
H_{\vc h, \mt J}(\vc s)=\sum_{i=1}^nh_is_i+\sum_{\langle i,j\rangle}J_{ij}s_is_j.
\end{equation}
Here, $\vc s$ is a vector of $n$ spins, $s_i\in\{-1, 1\}$, which carry an individual energy weight $h_i$ and are interconnected through 2-local couplings $J_{ij}$. The sum in the second term of the hamiltonian runs over only those spin pairs which are connected, as $J_{ij}=0$ for uncoupled spin pairs. As such, the hamiltonian is characterised by the linear weight vector $\vc h$ and the coupling matrix $\mt J$. The search for the minimum spin configuration $\vc s_{\rm min}$ for the Ising hamiltonian is known to be NP-hard \cite{Neukart2017b, Lucas2014}.\par
It is generally preferable for a computational application to work with $\{0, 1\}$-valued bits of information as opposed to spins, which can be achieved through the transformation $x_i=\frac12(s_i+1)$. This formulation of the Ising spin problem, which is polynomially reducible to the original form and vice versa, is known as a \textit{quadratic unconstrained binary optimization} problem, or QUBO for short, and can be solved by the QPU in the same fashion as a conventional Ising model. The equivalence between the two problem classes implies that any problem to be solved with the D-Wave QPU may be formulated either as a QUBO instance or directly as an Ising model. The objective quantity that the QPU minimizes in the QUBO case is given by the quadratic form \cite{Neukart2018, Neukart2017}
\begin{equation} 
\label{equation2}
\mathrm{Obj}_{\mt Q}(\vc x) = \vc x\T \mt Q\,\vc x,
\end{equation}
where $\vc x$ is an $N$-sized vector of $\{0, 1\}$-valued variables, and $\mt Q$ is an $N\times N$ real-valued upper triangular matrix containing the (adjusted) linear weights in the diagonal, and the couplings in the off-diagonal entries.\par
D-Wave's QPU physically implements an undirected graph in which the qubits, representing the binary variables, are the nodes, and the couplings the edges. The initial configuration is set up such that all  qubits are in uniform superposition, $\ket{x_i}=\frac1{\sqrt2}(\ket0+\ket1)$. During the annealing cycle, the state is evolved according to the energy landscape described by $\mt Q$. Eventually, when the system reaches the ground state, a minimum solution to the QUBO problem is found.\par
In this work, we demonstrate a method for using the QPU as an optimizer for a finite-element design problem. That is, we seek to optimize the shape of a 3D body defined by a finite number of elements against certain physical circumstances, by expressing the physical interaction of the elements in a QUBO form, and having the QPU find the minimum-energy configuration corresponding to a (sub)optimal shape. The next sections describe and discuss the problem in the framework of finite-element methods, as well as our approach to the problem and the observed results.\par
This paper is structured as follows. Sections 2 and 3 briefly discuss the research field of finite-element methods, the problem we address and its context in vehicle engineering, and how the two relate in our work. Section 4 outlines our method for solving the problem, including a detailed description of the QUBO formulation and the procedure that our proposed algorithm follows. Section 5 showcases the results in terms of shape optimization that we obtained by executing the algorithm, and examines a number of features and limitations of the algorithm that appear from these results. Lastly, we present our conclusions in section 6 and give an outlook on possible future work in section 7.
\section{Finite-element methods}
Finite-element methods (FEM) are a general group of numerical methods used in various physical tasks. Most well-known is the application of FEM in the investigation of the strength and deformation of solids with a geometrically complex shape, because here the use of classical methods, e.g. beam theory, proves to be too time-consuming or impossible. Logically, the FEM is based on the numerical solution of a complex system of differential equations. The computation domain, e.g. the solid, is divided into finitely many subdivisions of a simple form, creating a mesh of the original solid using a finite number of corner points (`vertices') and faces (`elements' or `simplices'). The physical behaviour of this meshed solid can be more easily calculated with well-known elementary functions due to its simplified geometry. The physical behaviour of the whole body is modelled in the transition from one element to the next, through very specific problem-dependent continuity conditions that must be fulfilled by the elementary functions. These functions contain parameters that usually have a physical meaning, such as the shift of a certain point in the component at a given time. The search for the motion function is thus returned to the search for the values functions' parameters. By using more and more parameters such as more and/or smaller elements and higher order functions, the accuracy of the approximate solution can be improved. The development of the FEM was possible in essential stages only by the development of powerful computers, since it requires considerable computing power. Therefore, this method was formulated from the outset to be processed on computers. Further information can be found in the work by Pepper et al. \cite{Pepper2017}.
\section{Quantum-assisted finite-element method for design optimization}
\begin{figure}[!b]
  \centering
  \begin{center}\includegraphics[scale=0.3]{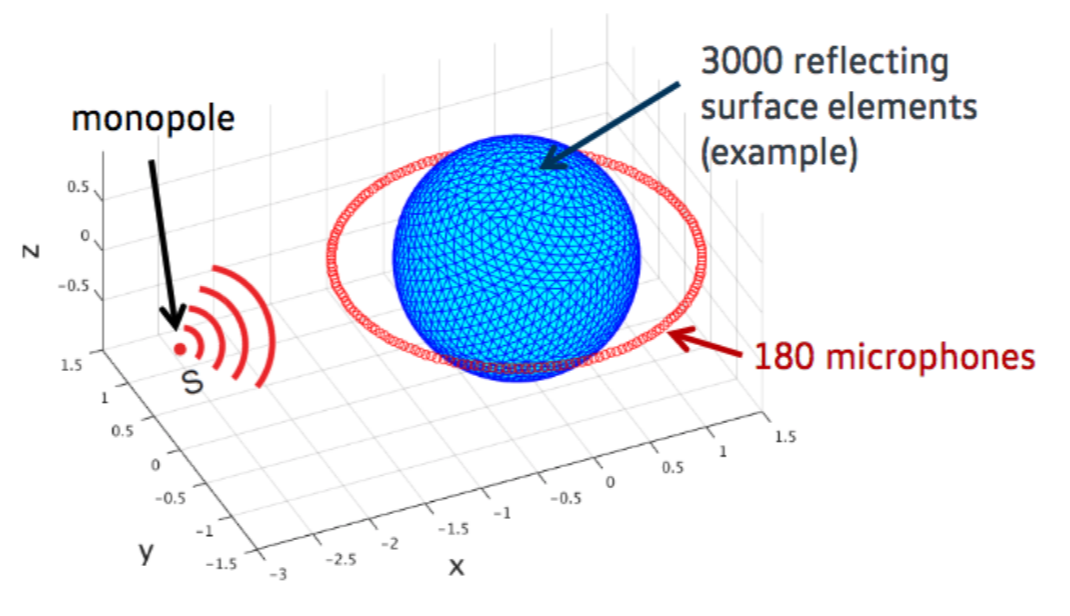}
  \end{center}%\rule[-.5cm]{0cm}{4cm} \rule[-.5cm]{0cm}{0cm}}
  \caption{Acoustic monopole emitting a spherical wave scattered by a rigid sphere.}
  \label{figure3}
\end{figure}
With the algorithm we introduce in the following sections, we are able to find designs based on a quantity that we minimize. One practical example concerns minimizing the wind noises on an external mirror of a vehicle, and another one is minimizing the noises through vibrations caused by the engine or different road conditions in a vehicle. The areas to optimize are commonly obtained with a complex finite-element simulation, and evolutionary algorithms have proven to be very valuable for searching the design space \cite{Sanz-Garcia2012, Back1997, Back2015, Duddeck2008}. As one part of the wind noise prediction simulation chain, we can compute acoustic sources on the mirror surface. This is an instance of a so-called acoustic scattering problem, which has to be solved in order to extract only those sources which are most relevant (noise-causing) at the position of the passengers. Solving the scattering problem is very time-consuming, especially in real vehicle applications, where the number of elements can be in the order of millions. Even for relatively few, a direct solver implementing straightforward matrix inversion quickly runs into memory and computation time limits. Thus, we are after finding an algorithm that scales better with an increasing number of elements. The present state of quantum computing does not allow us to compete with classical algorithms in terms of number of elements or speed, as the currently newest version of the QPU, containing approximately 2048 qubits, can only reliably find minor embeddings for shapes with up to 50 elements. Of course, we can split a QUBO instance with more than 50 elements and process problems of arbitrary size, but this significantly increases the required computation time.\par
In the introduced algorithm, we start with an initial shape and intend to find a shape that deflects sound waves emitted by an acoustic monopole source such that the sound pressure within an area at another position around the shape is minimized. In the same instance, our algorithm must be form-preserving, as in the end the shape should still resemble the initial design. In the scenario we describe, the initial shape is a spherical mesh with finitely many triangular surface elements (simplices), which is hit by sound waves emitted from an acoustic monopole (see fig. \ref{figure3}).
Fig. \ref{figure3} shows microphones positioned around the shape, and the objective is to minimize the sound pressure at any position of choice, by altering the sphere's shape. As the size of the current D-Wave QPU is limited to 2048 qubits and each qubit bears only 6 connections to neighboring qubits, we make a number of assumptions and approximations in order to make this problem feasible for submission to the QPU with a reasonable number of elements. More complex formulations however are possible, but adding more interactions would require more qubits, allowing processing of fewer elements within reasonable execution time.\\[0.6cm]~~
\section{Approach}
In the definition of the sound scattering problem, we make one major simplification to ensure that the resulting formulation is a finite-element method that is well-suited for the QPU, which is to approximate sound waves as rays (i.e. straight lines in 3D space). That is, propagation of sound waves is treated similarly to the propagation of light as done in graphical raytracing \cite{Appel1968, Whitted1980}, ignoring wave effects and interference altogether. The most important reason for this is that it allows us to consider each element separately in terms of its contribution to the measured sound pressure, as it avoids the necessity to construct a wave-based model harbouring high degrees of interaction between elements through sound wave interference. After all, the multitude of incident and scattered sound waves creates a highly complex situation that cannot be described without distant (i.e. non-neighbouring) element-element coupling. Since we seek to devise a QPU-assisted finite-element method for optimising a shape, finding a way to describe a `first-order approximation' with only neighbour couplings is more important than figuring out a very accurate scattering solution. Since we know that sound waves in reality reflect linearly off a surface identically to light rays, we use this as the approximation to base our quantum-assisted algorithm on.\par
\begin{figure}[!t]
  \centering
  \begin{center}\includegraphics[scale=0.37]{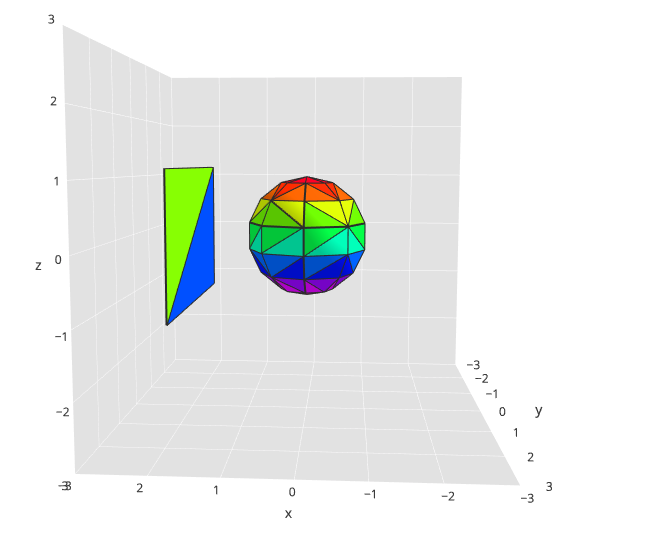}
  \end{center}%\rule[-.5cm]{0cm}{4cm} \rule[-.5cm]{0cm}{0cm}}
  \caption{A rigid sphere, which serves as the initial shape, and a rectangular area at which the sound pressure must be minimized. The mere purpose of the colour scheme is visual aid.}
  \label{fig:spherewithmic}
\end{figure}

The algorithm is a 3D search routine, which iteratively considers different candidate positions for each vertex in the shape (not to be confused with a qubit node in the QPU graph), and then lets the QPU decide which vertex arrangement causes the least number of rays to be reflected towards a microphone. This microphone is represented by a rectangularly bounded plane positioned next to the shape (see fig. \ref{fig:spherewithmic}). In each iteration, the routine assigns to each vertex $K$ `mutations', which are small random deviations from the original vertex position; that is, for each vertex $\vc v^i$ in the set $V$ of vertices, it considers $\vc v^i+\rmd\vc v^i_1, \cdots, \vc v^i+\rmd\vc v^i_K$ with $\rmd\vc v^i_j$ small. Each vertex is encoded by $K$ qubits, and the $\ket 1$ state of the $j$-th corresponding qubit indicates that mutation $j$ was assigned to this vertex (if the state is $\ket0$, this particular mutation was not assigned). For each simplex, the partial loss $\ell$, being the amount of pressure received from this simplex, is computed separately for each of the $K^3$ simplex configurations created from the vertex mutations (i.e. three vertices per simplex, and $K$ mutations for each vertex). The QUBO matrix $\mt Q$ is then constructed so that it contains, for each vertex, the loss information associated with the simplices neighbouring the vertex. Based on this information, the QPU will choose the minimal loss vertex configuration among the ones supplied, and use these as the input for the next iteration. This continues for a given number of iterations, or until convergence is observed.\par
A more detailed description of the QUBO formulation is provided in the next section.
\subsection{QUBO problem formulation}
Define $S$ as the set of all simplices $s$ determining the shape, $N=|V|$ and $C$ as the set of all configurations $c$ over the entire shape, where $c$ is a list of vertex mutation assignments $\{(i, j)\}$, with $i\in\{1, \ldots, N\}$ and $j\in\{1, \ldots, K\}$, indicating assignment of mutation $j$ to vertex $i$ (i.e. $\vc v^i\mapsto\vc v^i+\rmd\vc v^i_j$). Each configuration is a complete list, in that every vertex is assigned a mutation. Define a loss function $\mathcal L(S, c)$, which maps a configuration $c$ to a loss value, to be the total of the partial losses $\ell(s, c)$ of simplices $s\in S$ for configuration $c$,
\begin{align}
\mathcal L(S, c)=\sum_{s\in S}\ell(s, c),
\end{align}
and a loss partition function $\mathcal Z(S, C)$, the sum of the loss function over all configurations:
\begin{align}
\mathcal Z(S, C)=\sum_{c\in C}\mathcal L(S, c)=\sum_{c\in C}\sum_{s\in S}\ell(s, c).
\end{align}
In this form, $\mathcal Z$ is a function of $K^N$ configurations. Now, we observe that this sum can be rewritten by visiting all edges $(\vc v, \vc w)$ in the edge set $E$, and considering for each edge the two simplices adjacent to that edge. Since each simplex has three edges, this means each simplex is counted thrice, so we divide this new total by 3, to obtain:
\begin{align}
\mathcal Z(S, C)=\frac{1}{3}\sum_{c\in C}\,\sum_{(\vc v, \vc w)\in E}\,\sum_{s\in S_{(\vc v, \vc w)}}\ell(s, c),
\end{align}
where $S_{(\vc v, \vc w)}$ is the set of the two simplices adjacent to edge $(\vc v, \vc w)$.\par
Now notice that there are $K^{N-3}$ configurations which fix a triple of mutations for three vertices of a simplex $s$, and are thus equivalent for this particular simplex. As such, instead of counting each configuration separately, we consider only $K^3$ configurations that are nonequivalent with respect to this simplex to sum over (represented by the set $C_s$), and multiply the result by $K^{N-3}$:
\begin{align}
\mathcal Z(S, C)=\frac{K^{N-3}}{3}\sum_{(\vc v, \vc w)\in E}\,\sum_{s\in S_{(\vc v, \vc w)}}\,\sum_{c\in C_s}\ell(s, c).
\end{align}
This representation of the partition function now gives us an intuitive way to define a QUBO matrix $\mt Q$ for this problem. This instance is to be minimized by a $\{0, 1\}$-valued vector $\vc x$ representing a configuration $c(\vc x)$, whose entry corresponding to the mutation assignment $(i, j)$ is 1 if $\vc v^i$ is assigned mutation $j$ and 0 otherwise, as stated before. That is, we view each entry $x_{ij}$ as representing whether mutation $(i, j)$ is included in the configuration list of $c(\vc x)$ (in which case $x_{ij}=1$) or not (implying $x_{ij}=0$). The edge pairs naturally correspond to the off-diagonal terms of this matrix: for any edge pair $(\vc v^{i_1}, \vc v^{i_2})$ with mutations $(i_1, j_1)$ and $(i_2, j_2)$ respectively, we only need to sum over the partial loss values for all possible configurations regarding the two neighbouring simplices. If we define $\hat\ell(s, j_1, j_2, k)$ to be the partial loss from a simplex $s$ adjacent to edge $(\vc v^{i_1}, \vc v^{i_2})$ (that is, $s\in S_{(\vc v^{i_1}, \vc v^{i_2})}$) when its third, off-edge vertex is assigned mutation $k$ (while $\vc v^{i_1}$ is assigned mutation $j_1$ and $\vc v^{i_2}$ is assigned mutation $j_2$), we thus obtain the following matrix form:
\begin{align}
Q^{i_1j_1}_{i_2j_2}=\alpha\sum_{s\in S_{(\vc v^{i_1}, \vc v^{i_2})}}\,\sum_{k=1}^K\,\hat\ell(s, j_1, j_2, k).\label{vgl:qubo_entry}
\end{align}
Here, $\alpha$ is an energy scaling factor that absorbs the $K^{N-3}/3$ in front of the sum in eq. \ref{vgl:qubo_entry} (in practice, this $K^{N-3}$ will turn out to be huge, so adjustment is necessary). In this form of $\mt Q$, each entry fixes an edge, and a configuration for both vertices of this edge. Since $\mt Q$ contains $K$ rows and $K$ columns for each vertex, it is an $NK\times NK$ matrix.\par
Lastly, it is important to make sure the QPU returns a result vector $\vc x$ such that each vertex is being assigned only one mutation in the corresponding configuration $c(\vc x)$. Since $\vc x$ is binary, this is equivalent to requiring
\begin{align}
\forall i:\;0=\bigg(\sum_{j=1}^Kx_{ij}-1\bigg)^2=-\sum_{j=1}^Kx_{ij}+2\sum_{j=1}^K\sum_{j'>j}^Kx_{ij}x_{ij'}+1.
\end{align}
A straightforward way to enforce this requirement is by adding it as a penalty term to the loss function with some large constant penalty coefficient $\lambda$, as proposed in a recent paper on QPU traffic flow optimization \cite{Neukart2017b}:
\begin{align}
\tilde{\mathcal L}(S, \vc x)=\mathcal L(S, c(\vc x))+\lambda\sum_i\bigg(\sum_{j=1}^Kx_{ij}-1\bigg)^2.
\end{align}
In the QUBO matrix, this directly translates to adding $-\lambda$ to the diagonal elements $Q^{ij}_{ij}$ and adding $2\lambda$ to the off-diagonal elements $Q^{ij}_{ij'}$ ($j'>j$) corresponding to vertex $\vc v^i$. Provided $\lambda$ is large enough, this measure guarantees the QPU sets exactly one of the bits $x_{i1},\ldots, x_{iK}$ to 1, as any infeasible assignment would cause an increase in loss that would be higher than any possible gain from selecting a different configuration.
\subsection{Algorithm}
\begin{figure}[!b]
\centering
\includegraphics[width=0.85\textwidth]{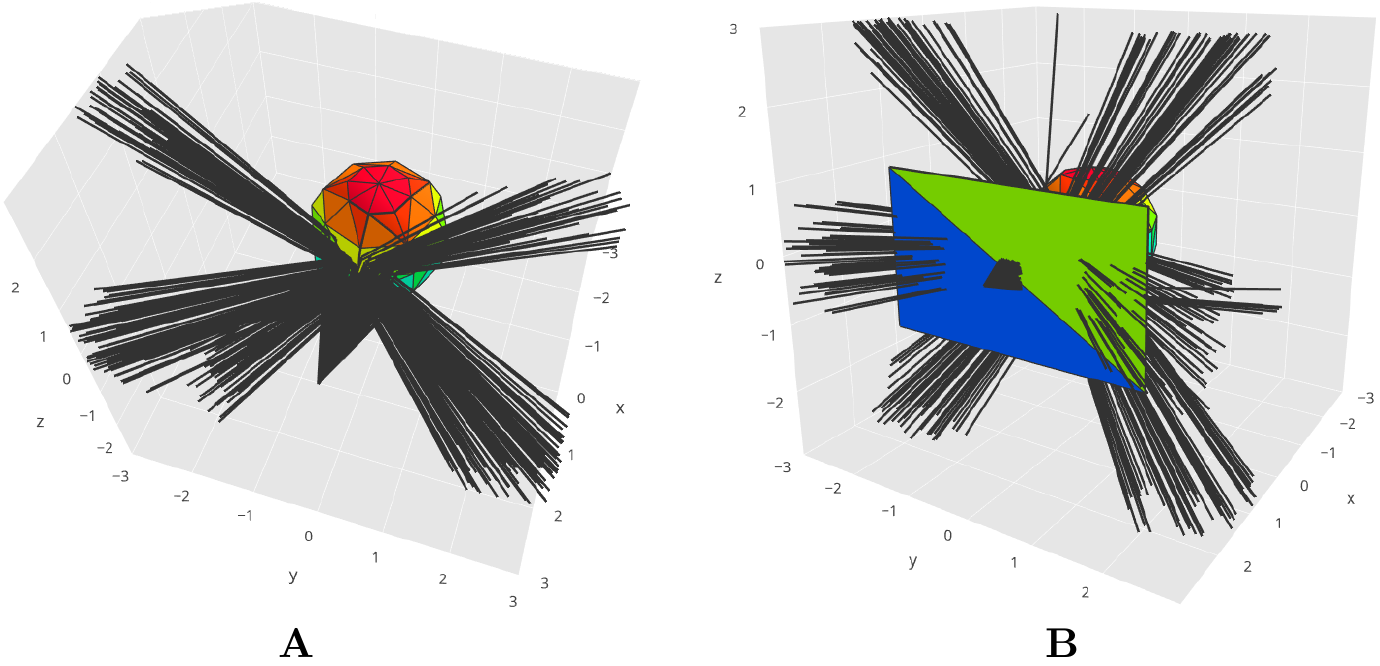}
\caption{Initial setup for the first experiment with the monopole at $(2.5, 0, 0)$. (\textbf{A}) Scattering of 300 incoming rays, casted from the monopole, off the spherical surface. (\textbf{B}) A significant number of rays is reflected backwards, intersecting the microphone. The rays crossing the microphone in the centre are considered incoming rays and are not counted towards the recorded sound pressure.}
\label{fig:init;M=[2.5, 0, 0]}
\end{figure}
With an overview of the procedure in our approach, and an explanation of the QUBO formulation, we can now turn to the algorithm itself. This iterative algorithm executes the following steps.
\begin{enumerate}[itemsep=0pt]
\item First, we generate the low-resolution mesh of an initial spherical shape. The vertices of this shape are conveniently represented as rectangular lattice points in the $(\theta, \phi)$ space of spherical coordinates (the radius $r$ may be chosen equal to unity without loss of generality). The edges of the mesh can then by found by Delaunay tessellation of this lattice. With the method of Delaunay triangulation, points in the $\mathbb R^2$ plane are transformed into triangles so that there are no other points within the circumscribed circle of each triangle. The method is used, for example, to optimize calculation networks for many finite-element methods. As a result, the triangles of the edge set have the largest possible internal angles; mathematically speaking, the smallest interior angle over all triangles is maximized. This feature is very desirable in computer graphics because it minimizes rounding errors. The algorithm responsible for computing Delaunay tessellations is explained in detail by Dobkin et al. \cite{Dobkin1996}. Given the vertices and edges in spherical coordinate space, a 3D spherical shape is constructed by the coordinate map $x_i=\sin\theta_i\cos\phi_i$, $y_i=\sin\theta_i\sin\phi_i$, $z_i=\cos\theta_i$. The convex hull of this shape is created around these 3D dots by drawing a face for all triangles, and the outward normal for each triangle is calculated. After this initial setup, the sequence of iterations starts.
\item As the first step in each iteration, $K$ vertex mutations are computed for each vertex. The mutations are chosen probabilistically such that $\rmd\vc v^i_j$ is within a sphere of decreasing radius $R_i=\beta\rho_it^{-\mu}$, with $t$ the current iteration and $\mu$ a constant. That is, each $\rmd\vc v^i_j$ is picked with (uniformly) random tangential and azimuthal angles, and uniformly random radius in the interval $[0, R_i)$. Here, $\rho_i$ is a shape-dependent bound for each vertex, whose purpose is to prevent the shape from becoming chaotic\footnote{By chaotic we mean the shape having too sharp corners, vertices extruding too far from the shape, edges intersecting other simplices etc., as well as the shape generally containing too many or too deep concavities. In practice, $\rho_i$ is determined by a soft convexity constraint which ensures that, as long as $\beta\leq1$, moving a vertex $\vc v_i$ by a distance $R_i$ in any direction will approximately retain the convexity of the shape. Since preserval of the convexity from the viewpoint of one vertex depends only on its neighbour vertices (and itself), $\rho_i$ is defined precisely by the position of $\vc v_i$ and the position of its neighbours.}. The factor $\beta$ acts as a control parameter setting the step size of the algorithm. Furthermore, in addition to this $(1, K)$-like search method (in analogy to $(1, \lambda)$ search in evolutionary strategies, with selection occuring in step 5), we implement an option for $(1+[K-1])$ search by allowing $\rmd\vc v^i_1=\vc0$ for all vertices $i$.
\item For each simplex $s$, we compute the $K^3$ partial loss values $\hat\ell(s, i, j, k)$. These are determined by casting a set number of rays towards that simplex when its first vertex is in mutation $i$, its second in mutation $j$ and its third in mutation $k$, and counting the fraction of rays that hits the rectangular microphone plane.
\item From these partial loss values, the $NK\times NK$-size QUBO matrix $\mt Q$ is computed as defined in the previous section. This matrix is then submitted to the QPU.
\item The QPU returns an $NK$-size bitstring $\vc x$ containing the preferred mutations of each vertex that yield minimal loss among all configurations. As mentioned in section 4.1, this bitstring is of the form $[x_{11}, x_{12}, \ldots, x_{1K};\;x_{21}, \ldots x_{2K};\;\ldots\;;$ $\;x_{N1}, \ldots, x_{NK}]$, where for each vertex $i$, only one of the bits $x_{i1}, \ldots, x_{iK}$ is 1, indicating the chosen preferred mutation for this vertex, and the others are 0. The shape is subsequently adapted according to this bitstring.
\item Steps 2--5 are repeated as often as necessary.
\end{enumerate}
In the following, we show and discuss some of the resulting shapes that we have obtained from running this algorithm.
\section{Experimental results and discussion}
\begin{figure}[!b]
\centering
\includegraphics[width=0.95\textwidth]{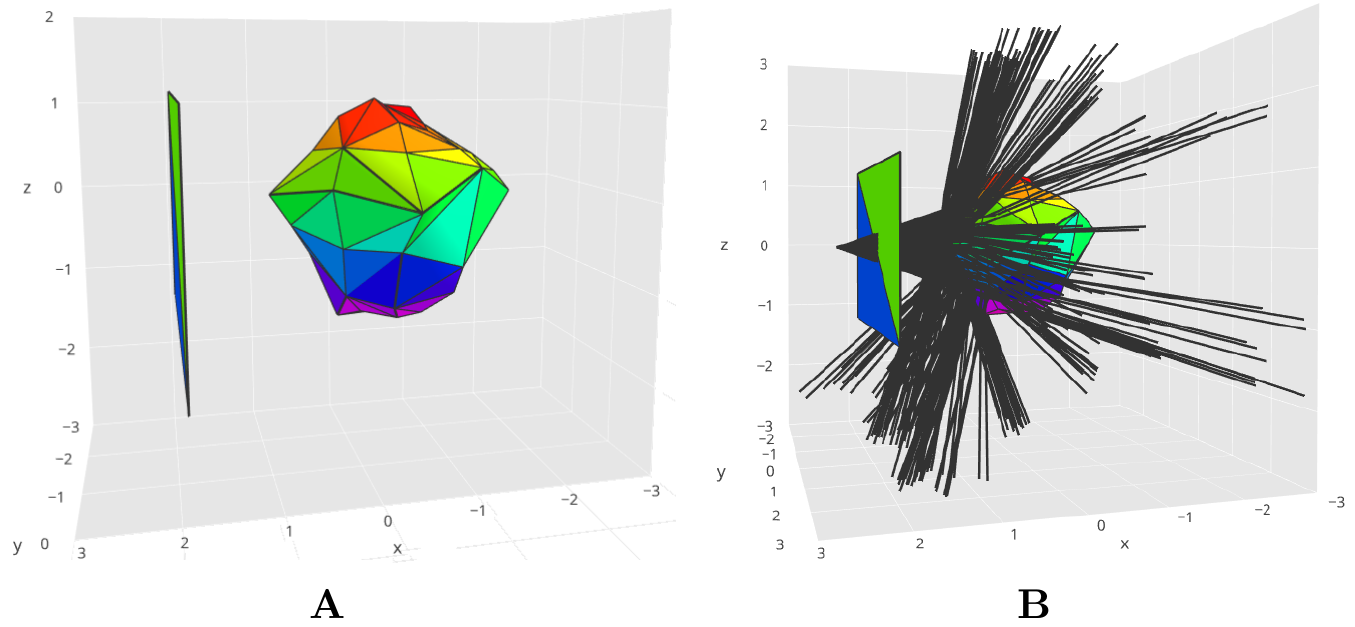}
\caption{Result after running the algorithm with the monopole at $(2.5, 0, 0)$ (see figure \ref{fig:init;M=[2.5, 0, 0]}). (\textbf{A}) The resulting shape. Again, rainbow colours are used for visual support. The shape displays a sharp edge at the front, giving it a streamlined structure. (\textbf{B}) Rays scattering off the new shape. All rays are directed around the microphone plane in one way or another, as can be seen from the fact that no outgoing ray intersects the microphone plane.}
\label{fig:opt;M=[2.5, 0, 0]}
\end{figure}
In our first experiment, we consider the situation with the monopole source sitting at $(2.5, 0, 0)$. The microphone is at $x\approx2$ and is approximately bounded by $y\in[-2, 2]$, $z\in[-1.15, 1.15]$. See figure
\ref{fig:init;M=[2.5, 0, 0]}. We run the algorithm with $K=3$, $\beta=0.7$ and $\mu=0.18$. At this point, we conduct $(1, K)$ search by having the routine choose $\rmd\vc v_0^i$ randomly, as described in section 4.2. For the computation of the partial loss values associated with the triangles, we sample 50 rays casted toward each triangle. It must be noted that often either all or none of the rays end up intersecting the microphone plane; however sampling more rays reduces potential variance in the partial loss calculations, making the algorithm more robust.\par
The resulting shape as determined by the algorithm is shown in figure \ref{fig:opt;M=[2.5, 0, 0]}. As one can see, the algorithm is successful in achieving its goal of minimizing the sound pressure, expressed in the amount of sound rays, at the microphone. It has found a way to adjust the front triangles such that each ray will either scatter in the negative $x$ direction or, if scattered backwards in the positive $x$ direction, travels around the microphone plane. This is clearly a consequence of the sharp tip the shape has obtained, which was absent in the case of the sphere.\par
\begin{figure}[!b]
\centering
\includegraphics[width=0.95\textwidth]{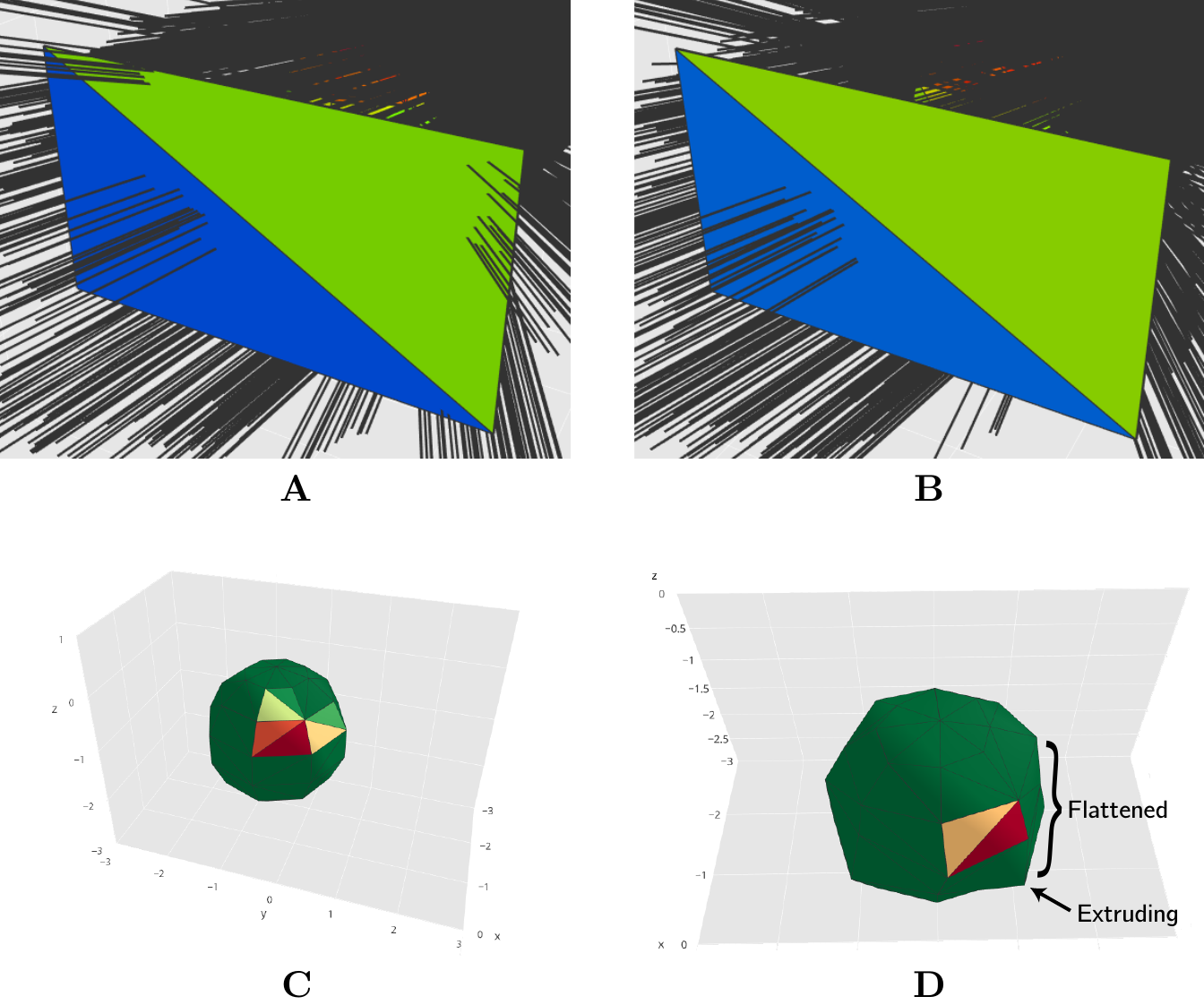}
\caption{Results and comparison after executing the algorithm with the monopole at $(0, 3, 2)$ with a lower step size. (\textbf{A}) At first, before shape adjustment, rays intersect the microphone at three positions: at the upper left, at the upper right and slightly to the left of the centre. (\textbf{B}) The shape returned by the optimisation algorithm now reflects the rays, which would initially hit the corners of the microphone, away from it. However, the centre rays seem to remain in place. (\textbf{C}) Partial loss shading of the initial sphere. Dark green triangles reflect no rays towards the microphone, while a darker shade of red indicates higher partial loss. Shade is normalised to the maximum partial loss of any triangle. (\textbf{D}) The final shape. One can see that the right side of the shape has been flattened, having an extruded point, which contributes to reflection away from the microphone.}
\label{fig:opt;M=[0, 3, 2]}
\end{figure}
It is worth noting that the rear of the sphere, at the far away end from the microphone, was deformed into a seemingly random structure. This is caused by the fact that no rays would hit this side in the first place; as such the quantum algorithm has no information about it (meaning the quadratic QUBO entries corresponding to those triangles are zero) and will choose a random vertex in each iteration. As such, it would make sense, in a further version of this algorithm, to prune these triangles in order to allow processing of more detailed shapes (containing more elements) on the QPU. In this work, however, we chose not to do this as our wish was to investigate the effect of the algorithm on the entire shape. After all, our problem was inspired by external vehicle mirror design, which does not allow for cut shapes.
The values for $\beta$ and $\mu$ were chosen by trial-and-error search, by testing a small set of combinations covering $\beta\in[0.3, 1.0]$ and $\mu\in[0.15, 0.20]$. We noticed that a too low step size renders the algorithm incapable of sufficiently adapting the shape within the given number of iterations, as it usually gets stuck in a local, suboptimal point, which cannot be optimized any further. This seems to occur in particular with $(1+[K-1])$ search. On the other hand, a too high step size usually (especially in the case of $(1, K)$ search) produces a too irregular shape. A good example showing the consequence of choosing a too low step size can be seen in figure \ref{fig:opt;M=[0, 3, 2]}. Here, we moved the monopole to $(0, 3, 2)$ and chose a step size control $\beta=0.3$. We observe that although two sources of loss have been eliminated, one seems to be persistent. The result in figure \ref{fig:opt;M=[0, 3, 2]}(d) with only two triangles having nonzero partial loss (which, even though not shown in the figure, is lower than that of the sphere in fig. \ref{fig:opt;M=[0, 3, 2]}(c)) is most likely considered as a local optimum by the algorithm, meaning it chooses not to depart from there.
\section{Conclusions}
In this work, we have presented a finite-element method for optimizing a three-dimensional shape under given physical criteria. By formulating an approximation of this finite-element problem in a QUBO form, and by embedding the corresponding matrix on the QPU as specified, we have been able to show that it is possible to successfully carry out finite-element design optimization on a D-Wave QPU. We have shown that by supplying an initial shape we can optimize the geometry to minimize a specified quantity, such as sound pressure, at a target area around the shape or the vibration of single elements, and in the same instance partially preserve the geometry. This is important, as if we supply the design of an outside mirror of a vehicle and intend to minimize the noise at the passenger's positions, we still want to end up with a design that captures all the properties a mirror must have.\par
Furthermore, we have demonstrated how to usefully combine the computing power of a classical computer with that of a quantum computer. That is, we calculate the sound pressure on an initial geometry classically and have the QPU solve the problem prepared on the classical computer. It is this combination of CPU and QPU efforts that in the end yields the desired solution.
\section{Future work}
For the next version of the algorithm, we intend to find a formulation that will incorporate additional constraints on the final shape. In addition, we would like to add wave behaviour corrections to increase the degree of realism in the model, or alternatively, discard the ray-casting approximation and find a way to model sound waves directly. Additionally, we wish to explore scalability of the algorithm, as we should be able to process shapes with more elements by splitting the QUBO matrix with the \texttt{qbsolv} decomposing solver tool \cite{D-Wavesystems2018}, instead of having the D-Wave software find minor embeddings for shapes with few elements. This will be of use in the future, when we expect new D-Wave QPU generations. With these new chips having more couplers between the qubits, we will be able to embed shapes with more elements and hopefully determine smoother geometries. We will continue to focus on laying the foundation for solving practically relevant problems by means of quantum computing, quantum simulation, and quantum optimization \cite{Neukart2013, Neukart2014, Eisenkramer2017, Neukart2017a, Levit2017, Crawford2016, Neukart2018, Neukart2017}.
\section*{Acknowledgments}
Thanks go to VW Group CIO Martin Hofmann and VW Group Region Americas CIO Abdallah Shanti, who enable our research.
\nocite{*}
\bibliographystyle{unsrt}
\bibliography{bibliography_acoustics}

\begin{thebibliography}{10}

\bibitem{Benedetti2016}
Marcello Benedetti, John Realpe-G{\'{o}}mez, Rupak Biswas, and Alejandro
  Perdomo-Ortiz.
\newblock {Estimation of effective temperatures in quantum annealers for
  sampling applications: A case study with possible applications in deep
  learning}.
\newblock {\em Physical Review A}, 94(2), 2016.

\bibitem{Smelyanskiy2017}
Vadim~N. Smelyanskiy, Davide Venturelli, Alejandro Perdomo-Ortiz, Sergey Knysh,
  and Mark~I. Dykman.
\newblock {Quantum Annealing via Environment-Mediated Quantum Diffusion}.
\newblock {\em Physical Review Letters}, 118(6), 2017.

\bibitem{Venturelli2015a}
Davide Venturelli, Dominic J.~J. Marchand, and Galo Rojo.
\newblock {Quantum Annealing Implementation of Job-Shop Scheduling}.
\newblock jun 2015.

\bibitem{Jiang2017}
Zhang Jiang and Eleanor~G. Rieffel.
\newblock {Non-commuting two-local Hamiltonians for quantum error suppression}.
\newblock {\em Quantum Information Processing}, 16(4), 2017.

\bibitem{Isakov2016}
Sergei~V. Isakov, Guglielmo Mazzola, Vadim~N. Smelyanskiy, Zhang Jiang, Sergio
  Boixo, Hartmut Neven, and Matthias Troyer.
\newblock {Understanding quantum tunneling through quantum Monte Carlo
  simulations}.
\newblock {\em Physical Review Letters}, 117(18), 2016.

\bibitem{OGorman2015}
B.~O'Gorman, R.~Babbush, A.~Perdomo-Ortiz, A.~Aspuru-Guzik, and V.~Smelyanskiy.
\newblock {Bayesian network structure learning using quantum annealing}.
\newblock {\em The European Physical Journal Special Topics}, 224(1):163--188,
  2015.

\bibitem{Rieffel2014}
Eleanor~G. Rieffel, Davide Venturelli, Bryan O'Gorman, Minh~B. Do, Elicia~M.
  Prystay, and Vadim~N. Smelyanskiy.
\newblock {A case study in programming a quantum annealer for hard operational
  planning problems}.
\newblock {\em Quantum Information Processing}, 14(1), 2014.

\bibitem{Venturelli2015}
Davide Venturelli, Salvatore Mandr{\`{a}}, Sergey Knysh, Bryan O'Gorman, Rupak
  Biswas, and Vadim Smelyanskiy.
\newblock {Quantum optimization of fully connected spin glasses}.
\newblock {\em Physical Review X}, 5(3), 2015.

\bibitem{Perdomo-Ortiz2015}
A.~Perdomo-Ortiz, J.~Fluegemann, S.~Narasimhan, R.~Biswas, and V.N.
  Smelyanskiy.
\newblock {A quantum annealing approach for fault detection and diagnosis of
  graph-based systems}.
\newblock {\em The European Physical Journal Special Topics}, 224(1):131--148,
  2015.

\bibitem{Boixo2014}
Sergio Boixo, Troels~F. R{\o}nnow, Sergei~V. Isakov, Zhihui Wang, David Wecker,
  Daniel~A. Lidar, John~M. Martinis, and Matthias Troyer.
\newblock {Evidence for quantum annealing with more than one hundred qubits}.
\newblock {\em Nature Physics}, 10(3):218--224, 2014.

\bibitem{Babbush2014}
Ryan Babbush, Alejandro Perdomo-Ortiz, Bryan O'Gorman, William Macready, and
  Alan Aspuru-Guzik.
\newblock {Construction of Energy Functions for Lattice Heteropolymer Models:
  Efficient Encodings for Constraint Satisfaction Programming and Quantum
  Annealing}.
\newblock In {\em Advances in Chemical Physics}, volume 155, pages 201--244.
  2014.

\bibitem{Smolin2014}
J.A. Smolin and G.~Smith.
\newblock {Classical signature of quantum annealing}.
\newblock {\em Frontiers in physics}, 2(52), 2014.

\bibitem{Perdomo-Ortiz2012}
Alejandro Perdomo-Ortiz, Neil Dickson, Marshall Drew-Brook, Geordie Rose, and
  Alan Aspuru-Guzik.
\newblock {Finding low-energy conformations of lattice protein models by
  quantum annealing}.
\newblock {\em Scientific Reports}, 2, 2012.

\bibitem{LosAlamos2016}
Los~Alamos national laboratory.
\newblock {D-Wave 2X quantum computer}, 2016.

\bibitem{Neukart2017b}
Florian Neukart, Gabriele Compostella, Christian Seidel, David von Dollen,
  Sheir Yarkoni, and Bob Parney.
\newblock {Traffic flow optimization using a quantum annealer}.
\newblock {\em Frontiers in ICT}, 4(29), 2017.

\bibitem{Neukart2018}
Florian Neukart, David {Von Dollen}, and Christian Seidel.
\newblock {Quantum-assisted cluster analysis}.
\newblock mar 2018.

\bibitem{Neukart2017}
F.~Neukart, C.~Seidel, G.~Compostella, and D.~{Von Dollen}.
\newblock {Quantum-enhanced reinforcement learning for finite-episode games
  with discrete state spaces}.
\newblock {\em Frontiers in physics}, 5(71), 2017.

\bibitem{Lucas2014}
A.~Lucas.
\newblock {Ising formulations of many NP problems}.
\newblock {\em Frontiers in physics}, 2(5), 2014.

\bibitem{Pepper2017}
D.W. Pepper and J.C. Heinrich.
\newblock {\em {The finite element method: basic concepts and applications with
  MATLAB, MAPLE and COMSOL}}.
\newblock CRC press, 3 edition, 2017.

\bibitem{Sanz-Garcia2012}
A.~Sanz-Garc{\'{i}}a, A.V. Pern{\'{i}}a-Espinoza,
  R.~Fern{\'{a}}ndez-Mart{\'{i}}nez, and F.J. Mart{\'{i}}nez-de
  Pis{\'{o}}n-Ascac{\'{i}}bar.
\newblock {Combining genetic algorithms and the finite element method to
  improve steel industrial processes}.
\newblock {\em Journal of Applied Logic}, 10(4):298--308, 2012.

\bibitem{Back1997}
T~B{\"{a}}ck, D~B Fogel, and Z~Michalewicz.
\newblock {Handbook of Evolutionary Computation}.
\newblock {\em Evolutionary Computation}, 2:1--11, 1997.

\bibitem{Back2015}
Thomas B{\"{a}}ck, Peter Krause, and Christophe Foussette.
\newblock {Automatic Metamodelling of CAE Simulation Models}.
\newblock {\em ATZ worldwide}, 117(5):36--41, 2015.

\bibitem{Duddeck2008}
Fabian Duddeck.
\newblock {Multidisciplinary optimization of car bodies}.
\newblock {\em Structural and Multidisciplinary Optimization}, 35(4):375--389,
  2008.

\bibitem{Appel1968}
Arthur Appel.
\newblock {Some techniques for shading machine renderings of solids}.
\newblock In {\em Proceedings of the April 30--May 2, 1968, spring joint
  computer conference on - AFIPS '68 (Spring)}, page~37, 1968.

\bibitem{Whitted1980}
Turner Whitted.
\newblock {An improved illumination model for shaded display}.
\newblock {\em Communications of the ACM}, 23(6):343--349, 1980.

\bibitem{Dobkin1996}
David~P. Dobkin, C.~Bradford Barber, and Hannu Huhdanpaa.
\newblock {The quickhull algorithm for convex hulls}.
\newblock {\em ACM Transactions on Mathematical Software}, 1996.

\bibitem{D-Wavesystems2018}
D-Wave systems.
\newblock {Qbsolv, a decomposing solver}, 2018.

\bibitem{Neukart2013}
Florian Neukart and Sorin-Aurel Moraru.
\newblock {On Quantum Computers and Artificial Neural Networks}.
\newblock {\em Signal Processing Research}, 2(1), 2013.

\bibitem{Neukart2014}
Florian Neukart and Sorin~Aurel Morar.
\newblock {Operations on quantum physical artificial neural structures}.
\newblock In {\em Procedia Engineering}, volume~69, pages 1509--1517, 2014.

\bibitem{Eisenkramer2017}
Sven Eisenkr{\"{a}}mer.
\newblock {Volkswagen trials quantum computers}, 2017.

\bibitem{Neukart2017a}
Florian Neukart.
\newblock {Quantum physics and the biological brain}.
\newblock In {\em Reverse Engineering the Mind}, pages 221--229. 2017.

\bibitem{Levit2017}
Anna Levit, Daniel Crawford, Navid Ghadermarzy, Jaspreet~S. Oberoi, Ehsan
  Zahedinejad, and Pooya Ronagh.
\newblock {Free energy-based reinforcement learning using a quantum processor}.
\newblock may 2017.

\bibitem{Crawford2016}
Daniel Crawford, Anna Levit, Navid Ghadermarzy, Jaspreet~S. Oberoi, and Pooya
  Ronagh.
\newblock {Reinforcement learning using quantum boltzmann machines}.
\newblock {\em arXiv preprint arXiv:1612.05695v2}, pages 1--17, 2016.

\bibitem{Back}
T.~B{\"{a}}ck and S.~Khuri.
\newblock {An evolutionary heuristic for the maximum independent set problem}.
\newblock In {\em Proceedings of the First IEEE Conference on Evolutionary
  Computation. IEEE World Congress on Computational Intelligence}, pages
  531--535. IEEE.

\bibitem{D-Wavesystems2017}
D-Wave systems.
\newblock {Quantum computing: how D-Wave systems work}, 2017.

\bibitem{Korenkevych}
D.~Korenkevych, Y.~Xue, Z.~Bian, F.~Chudak, W.~G. Macready, J.~Rolfe, and
  E.~Andriyash.
\newblock {Benchmarking quantum hardware for training of fully visible
  Boltzmann machines}.

\bibitem{Korenkevych2016}
Dmytro Korenkevych, Yanbo Xue, Zhengbing Bian, Fabian Chudak, William~G.
  Macready, Jason Rolfe, and Evgeny Andriyash.
\newblock {Benchmarking Quantum Hardware for Training of Fully Visible
  Boltzmann Machines}.
\newblock {\em Frontiers in physics}, 2(5), nov 2016.

\bibitem{Lanting2014}
T.~Lanting, A.~J. Przybysz, A.~Yu Smirnov, F.~M. Spedalieri, M.~H. Amin, A.~J.
  Berkley, R.~Harris, F.~Altomare, S.~Boixo, P.~Bunyk, N.~Dickson, C.~Enderud,
  J.~P. Hilton, E.~Hoskinson, M.~W. Johnson, E.~Ladizinsky, N.~Ladizinsky,
  R.~Neufeld, T.~Oh, I.~Perminov, C.~Rich, M.~C. Thom, E.~Tolkacheva,
  S.~Uchaikin, A.~B. Wilson, and G.~Rose.
\newblock {Entanglement in a quantum annealing processor}.
\newblock {\em Physical Review X}, 4(2), 2014.

\bibitem{Otterlo2012}
Martijn~Van Otterlo and Marco Wiering.
\newblock {Reinforcement learning and Markov decision processes}.
\newblock {\em Reinforcement Learning}, pages 3--42, 2012.

\bibitem{Sutton1998}
R.S. Sutton and A.G. Barto.
\newblock {\em {Reinforcement learning: an introduction}}.
\newblock MIT press, Cambridge, 1998.

\end{thebibliography}
\end{document}